\documentclass[journal=jacsat,manuscript=article, hyphens]{achemso}

\usepackage[version=3]{mhchem} 
\usepackage{url}
\usepackage{xcolor}
\usepackage{booktabs}
\usepackage{adjustbox,lipsum}
\usepackage{rotating}
\usepackage{wrapfig}
\usepackage{tablefootnote}
\usepackage{graphicx}   
\usepackage{subcaption} 
\usepackage{times}
\mciteErrorOnUnknownfalse

\author{Lukas Buecherl}
\affiliation[USU]{Biological Engineering Department, Utah State University, Logan, UT, USA}
\author{Felipe X. Buson}
\affiliation[UC]{School of Biological Sciences, University of Bristol, Bristol, UK}
\author{Georgie Hau Sørensen}
\affiliation[UB]{School of Biological Sciences, University of Bristol, Bristol, UK}
\author{Carolus Vitalis}
\affiliation[CU]{Department of Electrical, Computer, and Energy Engineering, University of Colorado Boulder, Boulder, CO, USA}
\author{Erik Kubaczka}
\affiliation[TUD]{Department of Electrical Engineering and Information Technology, Technical University of Darmstadt, Darmstadt, DE}
\author{Gonzalo Vidal}
\affiliation[CU]{Department of Electrical, Computer, and Energy Engineering, University of Colorado Boulder, Boulder, CO, USA}
\author{Bryan Bartley}
\affiliation[BBN]{Intelligent Software \& Systems, RTX BBN Technologies, Cambridge, MA, USA}
\author{Yan-Kay Ho}
\affiliation[UC]{Department of Chemical Engineering and Biotechnology, University of Cambridge, Cambridge, UK}
\author{G\"{o}ksel M{\i}s{\i}rl{\i}}
\affiliation[KU]{School of Computer Science and Mathematics, Keele University, Staffordshire, UK}
\author{Thomas E. Gorochowski}
\affiliation[UB]{School of Biological Sciences, University of Bristol, Bristol, UK}
\author{Jacob Beal}
\affiliation[BBN]{Intelligent Software \& Systems, RTX BBN Technologies, Cambridge, MA, USA}
\author{Chris J. Myers}
\affiliation[CU]{Department of Electrical, Computer, and Energy Engineering, University of Colorado Boulder, Boulder, CO, USA}
\author{Prashant Vaidyanathan}
\email{vprashant1@gmail.com}
\affiliation[OXB]{Data Science and Software Development, Oxford Biomedica, Oxford, UK}

\title{A decade of SBOL Visual: growing adoption of a diagram standard for engineering biology}

\abbreviations{SBOL}

\begin{document}

\begin{abstract}
\noindent
Standards play a crucial role in ensuring consistency, interoperability, and efficiency of communication across various disciplines. In the field of synthetic biology, the Synthetic Biology Open Language (SBOL) Visual standard was introduced in 2013 to establish a structured framework for visually representing genetic designs. Over the past decade, SBOL Visual has evolved from a simple set of 21 glyphs into a comprehensive diagrammatic language for biological designs. This perspective reflects on the first ten years of SBOL Visual, tracing its evolution from inception to version 3.0. We examine the standard’s adoption over time, highlighting its growing use in scientific publications, the development of supporting visualization tools, and ongoing efforts to enhance clarity and accessibility in communicating genetic design information. While trends in adoption show steady increases, achieving full compliance and use of best practices will require additional efforts. Looking ahead, the continued refinement of SBOL Visual and broader community engagement will be essential to ensuring its long-term value as the field of synthetic biology develops.
\end{abstract}

\section{Ten Years of SBOL Visual}

Standards play a crucial role in ensuring consistency, interoperability, and efficiency of communication across disciplines. They serve as a foundation for both interdisciplinary and international communities, enabling the seamless exchange of information by providing common frameworks for transferring information and understanding. Among these, visual standards are particularly important for communication and collaboration in technical and scientific fields, as they provide clear, structured, and universally understood representations of complex concepts. Well-known examples include the IEEE Standard Graphic Symbols for Logic Functions~\cite{IEEE1991Logic}, which defines standardized visual representations of logic gates, and the Unified Modeling Language (UML)~\cite{booch1997unified}, which provides a framework for visualizing system-level designs. While these standards apply to different domains, they share the overarching goal of providing a structured framework for creating figures and diagrams that effectively communicate complex domain specific information.

As the field of synthetic biology has expanded, the need for more standardized approaches to visualization has become apparent \cite{sbolVCommunication}. This led to the development of the Synthetic Biology Open Language (SBOL) Visual standard in 2013, to provide a standardized framework for representing genetic design information \cite{SBOLvisual1}. A decade later, we now evaluate the first ten years of SBOL Visual, trace its adoption over time, and look to its future. 

Like other visual standards, SBOL Visual has a simple mission: to enhance the clarity of diagrams and figures, particularly focusing on those capturing genetic design information. It serves as a powerful tool for communicating the structural details of nucleic acid sequences and the functional relationships between their features and other molecular species. Prior to its inception, the synthetic biology community relied on a vague consensus for visualizing such systems. The SBOL Visual specification aimed to consolidate, organize, and systematize these common practices into a cohesive visual language for genetic design.

\begin{figure}[h]
\centering
\includegraphics[width=8cm]{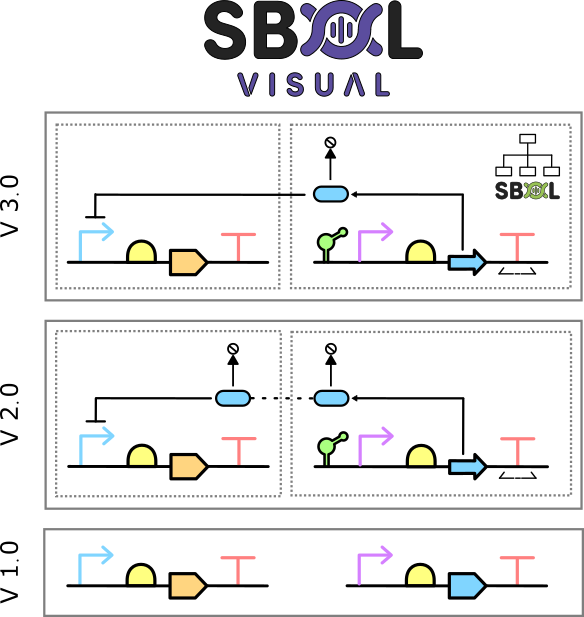}
\caption{Development of the SBOL Visual standard over the past decade. The initial release focused on the representation of functional information related to nucleic acid sequences. Version 2.0 not only expanded the glyph library, but also introduced representations for functional interactions and other molecular species. Finally, version 3.0 saw the seamless integration of the visual standard with the SBOL data model.}
\label{fig:SBOLVdevelopment}
\end{figure}

Initially, SBOL Visual was introduced as a collection of glyphs designed to represent common sequence features. Over the past decade, it has evolved into a comprehensive diagram language, expanding both its capabilities and leading to its widespread adoption. As shown in \textbf{Figure~\ref{fig:SBOLVdevelopment}}, this development has been marked by three key milestones which have  guided SBOL Visual from its initial release as version 1.0~\cite{SBOLvisual1} to its current iteration, version 3.0~\cite{SBOLvisual3, sbol31, sbol3spec}, with each major version reflecting the ongoing refinement of the related SBOL data model (an aligned effort to capture genetic design data across scales in a structured and machine-readable format)~\cite{SBOL3, sbolWorkflows}.

The first version of SBOL Visual~\cite{SBOLvisual1} was composed of a set of 21 distinct glyphs, designed to represent the functional elements of nucleic acid sequences. These included commonly used components such as promoters, ribosome entry/binding sites, protein coding sequences, and terminators. These glyphs purposefully avoided specifying stylistic features like line width or color, focusing instead on distinctive shapes, display names, and definitions. The definition of each glyph is established through its association with corresponding terms in the Sequence Ontology (SO)~\cite{eilbeck_sequence_2005}, which also helps to tightly align the visual standard with the SBOL data model~\cite{SBOL3} and support integration into broader design workflows~\cite{sbolWorkflows}.

With the release of version 2.0~\cite{SBOLvisual2, sbolVisual21, sbolVisual22}, SBOL Visual expanded its scope significantly, introducing 17 new glyphs. This version also introduced families of glyph variants (families of related glyphs that adhere to a coherent shape), enhancing the flexibility of the visual language. Furthermore, this version established an explicit connection between SBOL diagrams and the SBOL data model version 2.0~\cite{sbol23}, ensuring a more structured representation of genetic designs, as well as incorporating representations of various functional interactions and molecular species. This allowed SBOL Visual to more comprehensive capture the full range of components making up a genetic design~\cite{sbolVCommunication}.

The most recent iteration of SBOL Visual, version 3.0~\cite{SBOLvisual3}, was published in 2021 and further enhanced the standard by integrating it with the SBOL data model version 3.0~\cite{SBOL3,sbol3spec,sbol31}. While this version included several other minor updates, one notable change was the removal of dashed undirected lines used for subsystem mappings, as illustrated in \textbf{Figure~\ref{fig:SBOLVdevelopment}}. The standard was also expanded to represent a wider variety of interaction types, including cases where molecules inhibit or activate other interactions. These changes helped to improve the clarity and consistency of diagrams, aligning it more closely with the evolving data model.

\begin{figure}[p]
\centering
\includegraphics[width=8cm]{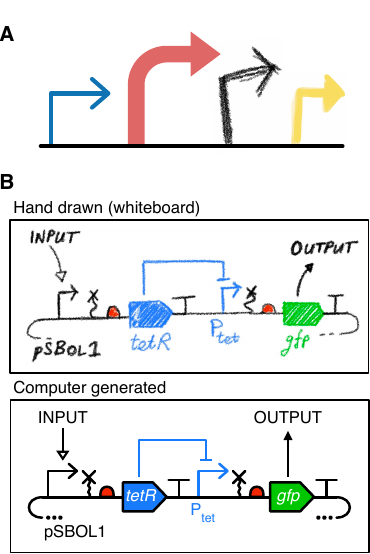}
\caption{SBOL Visual does not impose constraints on how diagrams are generated. (\textsf{\textbf{A}}) Differently styled versions of the same promoter part showing how the clear geometries of each glyph helps to aid interpretation. Users have the option to draw glyphs by hand, using computational tools, and customize their general aesthetic as long as the core shape of the glyph is legible. All examples shown would be compliant promoter glyphs. (\textsf{\textbf{B}}) Two identical genetic designs, one drawn by hand (top) and the other using a computational design software (bottom). Both clearly communicate the core design features.}
\label{fig:SBOLVstyles}
\end{figure}

Throughout the development of SBOL Visual, the primary goal has been to ensure simplicity, ease of use, and flexibility, while enabling the creation of clear and effective diagrams, both by hand and using computational tools (\textbf{Figure~\ref{fig:SBOLVstyles}}). To aid adoption, all SBOL Visual glyphs are publicly available in standard image formats on the SBOL website and can also be accessed via the projects GitHub repository (\textbf{Data Availability}). These glyphs can be used with any general-purpose computational drawing tool. 
Additionally, several specialized tools have been developed specifically for creating SBOL Visual diagrams, such as SBOL Canvas~\cite{terry2021sbolcanvas}, VisBOL~\cite{mclaughlin2016visbol}, SBOL Designer~\cite{zhang2017sboldesigner}, DNAplotlib~\cite{der2017dnaplotlib}, and paraSBOLv~\cite{clark2021parasbolv}.

Beyond the standard's use in creating static diagrams, SBOL Visual glyphs are also machine-accessible via the SBOL Visual Ontology, allowing computational tools to search for and interpret glyphs programmatically~\cite{sbolvisualontology}. The Visual Ontology Web Service enables the retrieval of specific glyphs based on terms from SO and the Systems Biology Ontology, providing an automated mapping service for users. Furthermore, ACS Synthetic Biology has officially endorsed the use of SBOL Visual, further solidifying its role as a recognized standard in the field~\cite{acsadoption}.

\section{Increasing Adoption over Time}

Given that over a decade has now passed since SBOL Visual was first released, we wanted to assess its adoption and impact across the scientific community. To quantify its use, we collected statistics on all SBOL Visual-compatible diagrams from the ACS Synthetic Biology journal in volumes published between 2012 and 2023. Notably, this dataset included 2012, which was when the inaugural issue of ACS Synthetic Biology was published, but a year before the release of SBOL Visual.

To assess adoption, we developed a method for classifying the compliance of a figure from a publication to the SBOL Visual standard. Figures were manually extracted from each publication and a team of reviewers then conducted the analysis, with each reviewer assigned figures from a separate calendar year to evaluate. For each extracted figure, an initial assessment determined whether SBOL Visual was relevant for its representation. For figures depicting genetic diagrams that could be represented using SBOL Visual, a reviewer then manually assessed compliance with the SBOL Visual Version 3.0 specification. Instances of non-compliance were documented to enable further analysis of general trends in compliance. If a figure was found to be SBOL Visual-compliant, the reviewer performed a further evaluation to assess whether it adhered to best practices outlined in the SBOL Visual specification. A figure was considered SBOL Visual-compliant if it adhered to all mandatory design rules outlined in the SBOL Visual specification, defined by the terms ``MUST'' and ``MUST NOT'' (\textbf{Figure~\ref{fig:SBOLVcompliance}}). In contrast, a figure was said to follow best practices if it complied with all recommended design rules, specified by the terms ``SHOULD'' and ``SHOULD NOT''  (\textbf{Figure~\ref{fig:SBOLVbestpractices}}). A comprehensive overview of all rules is provided in \textbf{Figure~\ref{fig:SBOLVcompliance}} and \textbf{Figure~\ref{fig:SBOLVbestpractices}}. In cases where the compliance or best practice of a figure was ambiguous, the reviewer shared the figure with the review team and a consensus decision made. These discussions ensured consistency in how figures were assessed across reviewers. In addition to compliance evaluations, group discussions served as a forum for identifying trends in non-compliance across different publications. Based on these discussions, a small number of exceptions to the SBOL Visual specification were agreed upon, as certain recurring cases were commonly seen in the analyzed publications. These special cases included: (1) using arrow glyphs to represent a coding sequence is accepted as best practice; (2) using a straight line to indicate restriction site is accepted as best practice; and (3) using a double slash line break for omitted material is compliant but not best practices.

Using this approach, we found a steady increase in the use of SBOL Visual-compliant diagrams, with the percentage of compliance approximately doubling over a decade. \textbf{Figure~\ref{fig:SBOLVadoption}} shows the percentage of SBOL Visual compliant figures per year, separating those that adherence to mandatory rules (i.e., compliance, in blue) and recommended guidelines (i.e., best practice, in orange). It should be noted that we only considered figures that could be represented as a genetic design where SBOL Visual is relevant. A spike in figures that were compliant and following best practices was seen in 2013, the year SBOL Visual was introduced, and then over the subsequent decade adoption has gradually risen with $>$70\% of genetic designs being SBOL Visual compliant since 2020. These results are incredibly promising. However, diagrams that fully adhered to best practices were fewer in number, being approximately 40\% lower than for general compliance. We also found that many diagrams continue to exhibit inconsistencies, such as the use of generic glyphs instead of standard-specific ones or the misapplication of glyphs for molecular features. This highlights an ongoing need for education and training to support researchers in applying SBOL Visual standards consistently across their diagrams. Importantly though, there were also growing numbers of recent examples that fully adhered to best practices (e.g., see \textbf{Figure~\ref{fig:SBOLVexamples}}). 

The steady adoption of SBOL Visual that we see can likely be attributed to ongoing community-driven efforts. These include the organization of workshops at events like the International Workshop on BioDesign Automation (IWBDA), and the COmputational Modeling in BIology NEtwork (COMBINE) workshops~\cite{IWBDA, myers_brief_2017}, publications in key scientific journals~\cite{sbolWorkflows, sbolVCommunication}, and the development of dedicated software tools to simplify the creation of SBOL Visual-compliant diagrams. Specifically, the development of the aforementioned specialized tools has contributed to increased adoption of the standard.  
It should be noted though that effective use of these tools often depends on a solid understanding of the SBOL Visual standard itself. This means that users less familiar with the standard may find it challenging to fully leverage the features available, making this another area that requires an investment in training and support.

\section{Outlook and Future Directions}

Our analysis has demonstrated an increasing adoption of SBOL Visual within the synthetic biology community. However, there is still room for improvement as the number of diagrams fully adhering to best practices significantly lags behind those that are compliant. To understand where to focus our efforts, it is important to go beyond the quantitative uptick in adoption and look for trends in the specifics of adoption over time. We identified several key types of non-compliance that can be targeted, and propose initiatives from the community of maintainers of the standard, as well as the wider synthetic biology community, to increase SBOL Visual uptake.

Many cases of non-compliance arose from simple factors, such as authors likely being unaware of the standard or using software tools that by default generate non-compliant nucleic acid diagrams. SBOL Visual was originally designed to incorporate existing glyphs already used by the community wherever possible, and as a result, it is unclear how much of the observed increase in compliance is driven by growing awareness of SBOL Visual itself, versus authors simply replicating genetic diagram formats they frequently encounter in recent publications. The most common faults in this category were the use of incorrect symbols for elements that have an associated SBOL Visual glyph (\textbf{Figure~\ref{fig:SBOLVcompliance}}, compliance rule 5.2.6, for example, using ovals for RBSs or stop signs for terminators) or overuse of the arrow glyph to represent a generic ``region with directionality'', being applied to a variety of elements including promoters, coding sequences, whole genes, origins of replication, and terminators. Furthermore, interactions were sometimes indicated solely by proximity rather than using arrows, leading to potential ambiguity (\textbf{Figure~\ref{fig:SBOLVcompliance}}, compliance rule 5.3.2). 

Another common occurrence that is compliant, but which deviates from best practices, was the use of generic glyphs instead of the more specific ones that exist. For example, rectangles (generic region) are used to represent elements of a known function (\textbf{Figure~\ref{fig:SBOLVbestpractices}}, best practice 5.2.6). Moreover, certain designs depicted both single and double-stranded DNA as a single backbone within the same diagram (\textbf{Figure~\ref{fig:SBOLVbestpractices}}, best practice 5.1.1), leading to inconsistency and possible confusion. Lastly, in some instances, designs did not provide a clear distinction between the backbone and its associated features.

Other cases of non-compliance showed that there was some knowledge of the standard, but that its application in complex contexts was difficult to understand. This occured in specialized fields such as mammalian synthetic biology and CRISPR system engineering, where existing SBOL Visual glyphs are sufficient for design diagrams, yet researchers opt for new symbols they consider more precise for the specific context they are considering. For example, we observed a trend in which a small black diamond is used as a spacer in CRISPR systems, and a double slash represents an indefinite number of bases between two designed sections within the same sequence. These new glyphs often appear to be informally adopted and passed down from paper to paper and it has been proposed that we should consider adopting these naturally occurring consensuses in future versions of SBOL Visual, since they are useful representations that do not clash with the current definitions.

There were also cases where SBOL Visual was simply not designed to represent the intricacies of the system being designed. Key examples include RNA and protein molecule design, representations that zoom into the bases of a DNA strand, and more abstract diagrams representing ideas such as workflows. In such cases, it will be necessary to consider significant revisions or expansions of the standard.

Other than improving the standard itself, increasing SBOL Visual's adoption will require additional efforts by the community to make it more accessible and known. To support this, maintaining and enhancing the tools that support diagram creation will be essential. This includes ensuring the latest glyph library is distributed on the SBOL website and via compatible tools such as SBOLCanvas~\cite{terry2021sbolcanvas}, DNAplotlib~\cite{der2017dnaplotlib}, and paraSBOLv~\cite{clark2021parasbolv}. Some new ways of generating genetic designs have also been proposed, such as a font that would allow the ``writing'' of constructs using any modern text editor. More efforts are also likely needed to further communication of the standard, including the rules for compliance and our best practices. This could be supported by additional workshops, hands-on tutorials and protocols~\cite{Bartoli2018}, and more detailed resources on the SBOL website and building closer connections with the wider computational biology communities (e.g., COMBINE)~\cite{COMBINEstandards}.

We envision that initiatives led by the wider synthetic biology community can also help accelerate uptake of SBOL Visual in a number of ways. While authors ultimately decide whether to use the standard or not, journal editors and reviewers also play a crucial role in identifying and addressing unclear or ambiguous figures and could suggest the use of SBOL Visual to improve the clarity of published research. Furthermore, companies and organizations could consider the use of SBOL Visual in their communications to more clearly convey their engineered biological designs, and developers of computational biodesign tools could broaden exposure of SBOL Visual by adopting the standard for visualization needs. Building links to these groups will be essential, and will likely require targeted efforts to ensure key individuals and organizations buy into using the standard more widely.

It is extremely promising to see that after a decade of SBOL Visual the communication of biological designs is clearer than ever. Now that standard has solidified and development of extensions and refinements slowed, it is crucial that the supporting infrastructure to enable even more widespread use is put in place. Here we have seen that through some specific targeted efforts the outlook is bright for unambiguous communication of engineered biology through diagrams.

\section{Data Availability}

SBOL Visual glyphs are available from the project website: \url{https://sbolstandard.org/visual-about/}. They can also be accessed via the SBOL Visual GitHub repository in both Portable Network Graphic (PNG) and Scalable Vector Graphics (SVG) formats: \url{https://github.com/SynBioDex/SBOL-visual/tree/master/Glyphs}. Detailed results for each paper and figure analyzed are available at \url{https://sbolstandard.org/sbolv-10-years/} and are also provided as supplemental spreadsheets accompanying this manuscript. Furthermore, we invite the community to provide feedback and get involved through the SBOL Enhancement Proposal (SEP) process to ensure that the standard continues to evolve with the needs of designers. Proposed ideas and issues can be submitted on GitHub: \url{https://github.com/SynBioDex/SBOL-visual/issues}.

\section{Author Contributions}

All authors contributed to the data collection and the writing of the paper. The authors created all figures in their entirety and all images used in the TOC graphic, except for Figure~\ref{fig:SBOLVexamples}. L.B. presented this work at 16th International Workshop on BioDesign Automation.

\begin{acknowledgement}

We would like to thank all contributors to the SBOL Visual standard over the past decade and members of the wider community. E.K. was supported by ERC-PoC grant PLATE (101082333). T.E.G. was supported by a Royal Society University Research Fellowship grant URF/R/221008, a Turing Fellowship from The Alan Turing Institute under EPSRC grant EP/N510129/1, and the UKRI-funded Engineering Biology Mission Award CYBER under BBSRC grant BB/Y007638/1. Y.K.H. and F.X.B. were supported by the UKRI BBSRC grant BB/W014173/1. C.V., G.V., and C.J.M. were supported by the Army Research Office under Cooperative Agreement Number W911NF-22-2-0210 and by DARPA grant number HR0011-24-C-0423. G.M. was supported by the BBSRC grant BB/Z517367/1. The funders had no role in study design, data collection and analysis, and the decision to publish or preparation of the manuscript. Any opinions, findings, and conclusions or recommendations expressed in this material are those of the author(s) and do not necessarily reflect the views of the funding agencies.
This document does not contain technology or technical data controlled under either U.S. International Traffic in Arms Regulation or U.S. Export Administration Regulations. 

\end{acknowledgement}

\bibliography{sbolvisual_at_ten}

\newpage
\section{Figures and captions}

\clearpage

\clearpage
\begin{figure}[p]
    \centering
    \includegraphics[width=1\linewidth]{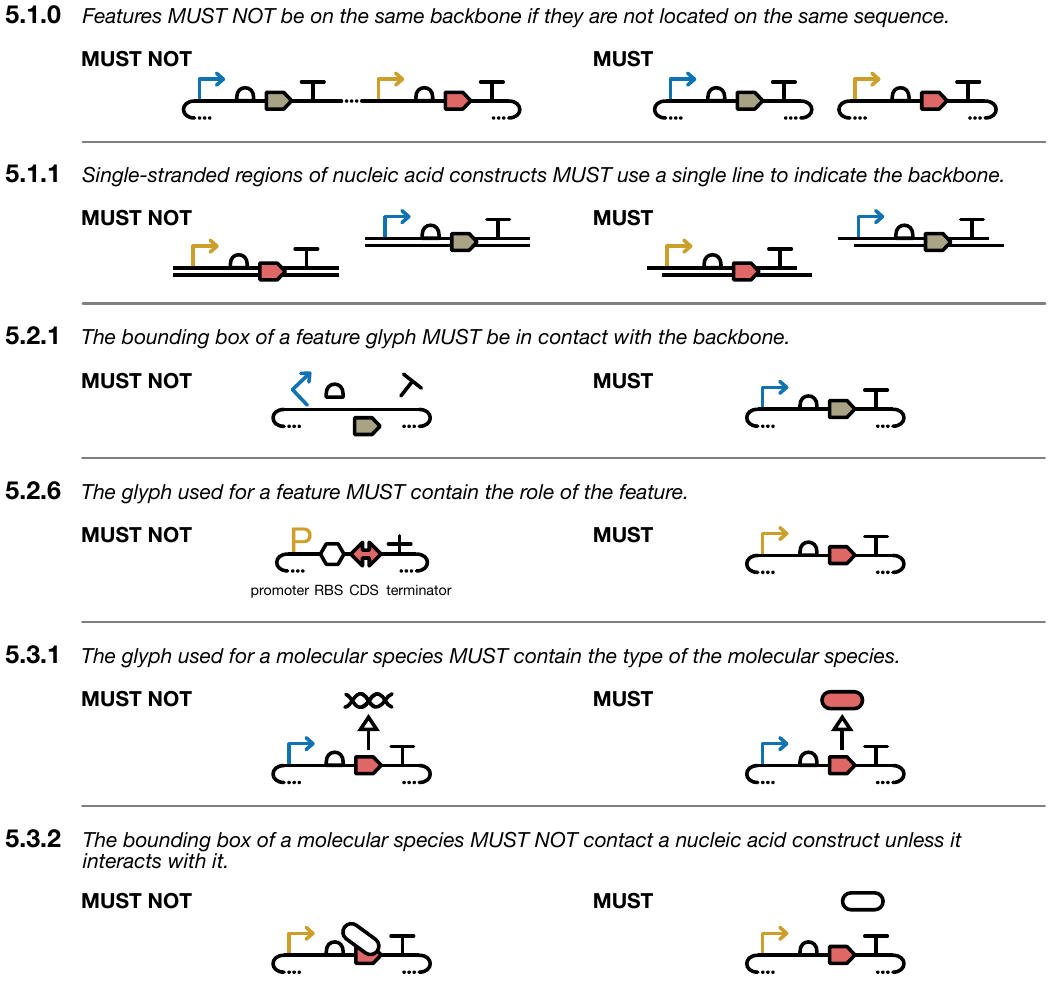}
    \caption{Examples of key SBOL Visual compliance rules.}
    \label{fig:SBOLVcompliance}
\end{figure}
\addtocounter{figure}{-1} 
\clearpage
\begin{figure}[p]
    \centering
    \includegraphics[width=1\linewidth]{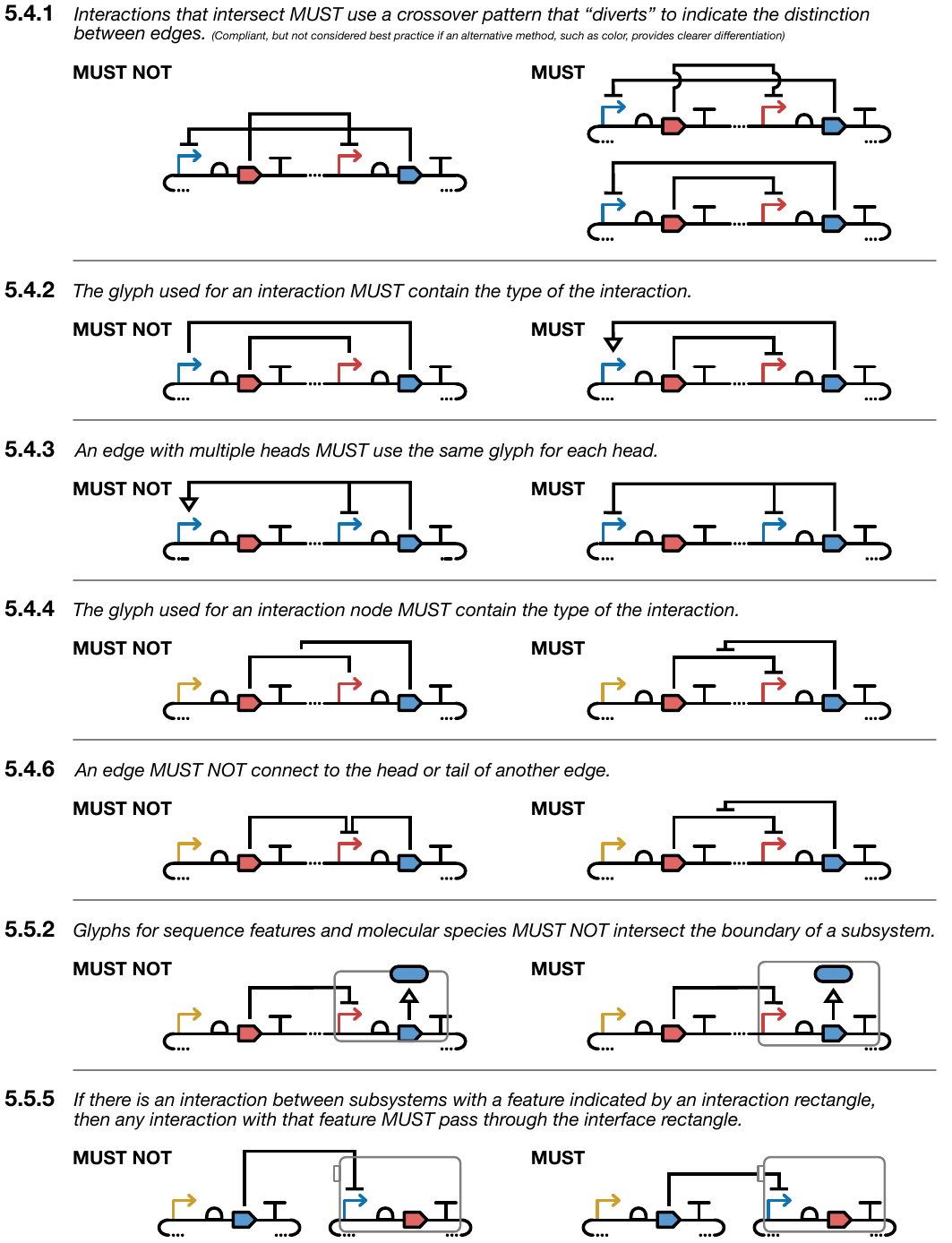}
    \caption{Examples of key SBOL Visual compliance rules.}
    \label{fig:SBOLVcompliance}
\end{figure}

\clearpage
\begin{figure}[p]
    \centering
    \includegraphics[width=1\linewidth]{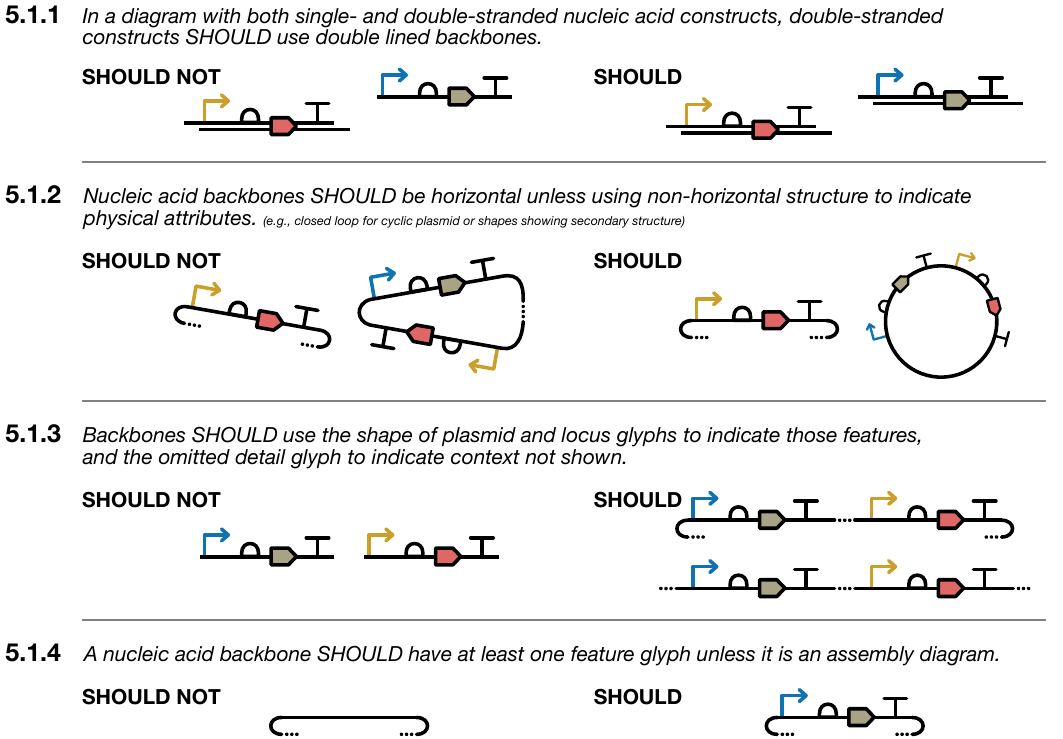}
    \caption{Examples of SBOL Visual best practices.}
    \label{fig:SBOLVbestpractices}
\end{figure}
\addtocounter{figure}{-1} 
\clearpage
\begin{figure}[p]
    \centering
    \includegraphics[width=1\linewidth]{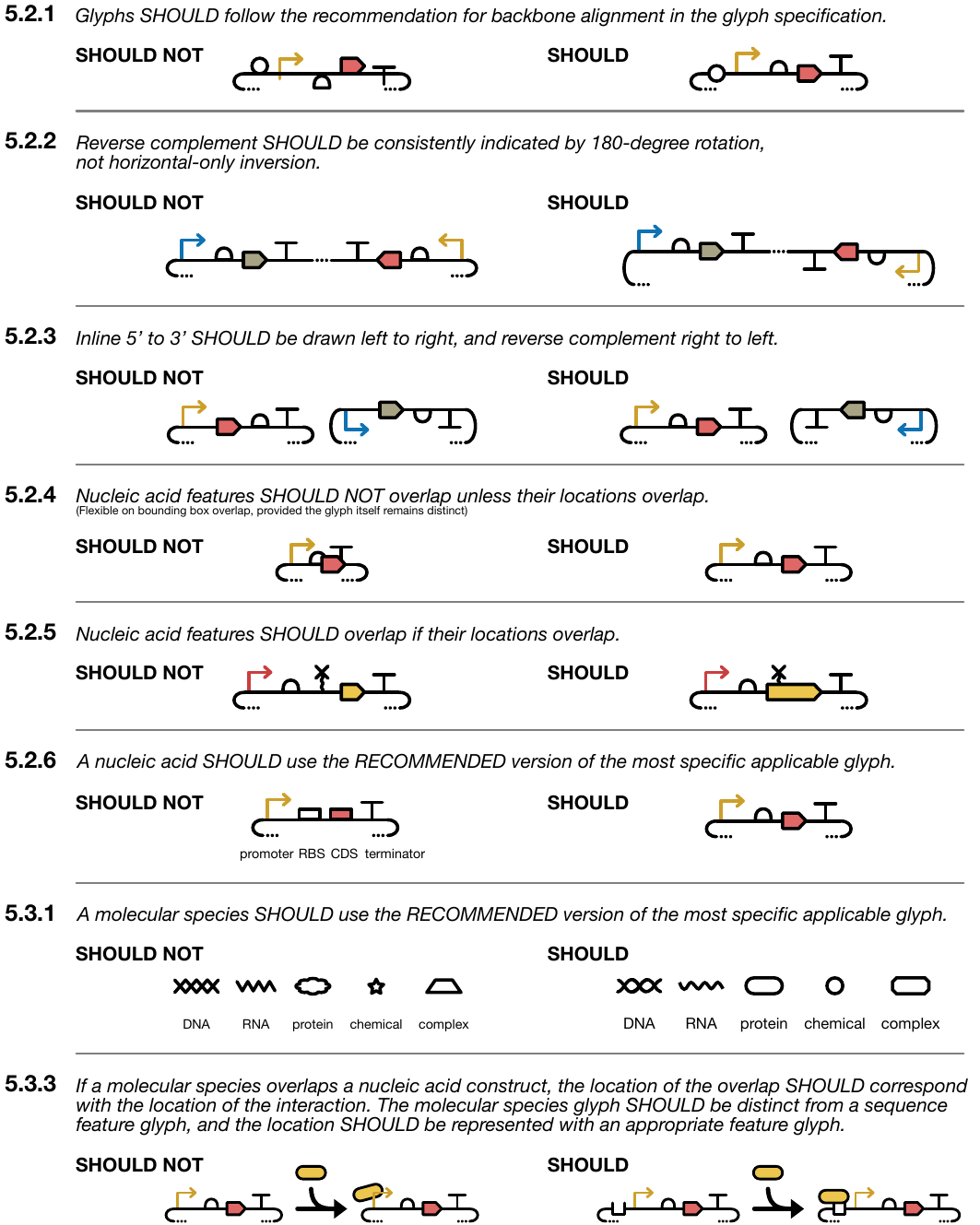}
    \caption{Examples of SBOL Visual best practices.}
    \label{fig:SBOLVbestpractices}
\end{figure}
\addtocounter{figure}{-1} 
\clearpage
\begin{figure}[p]
    \centering
    \includegraphics[width=1\linewidth]{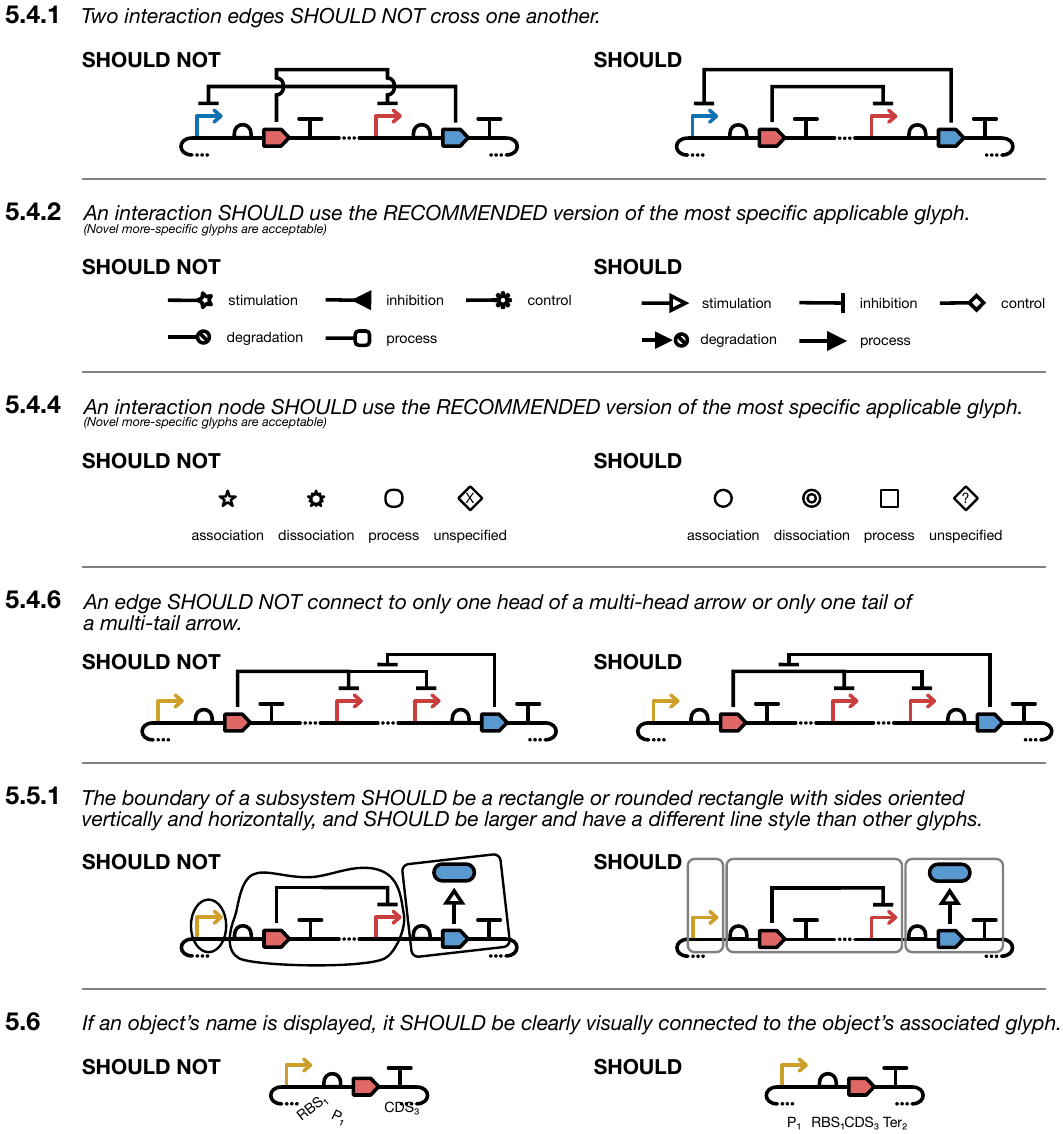}
    \caption{Examples of SBOL Visual best practices.}
    \label{fig:SBOLVbestpractices}
\end{figure}

\clearpage
\begin{figure}[p]
    \centering
    \includegraphics[width=0.7\linewidth]{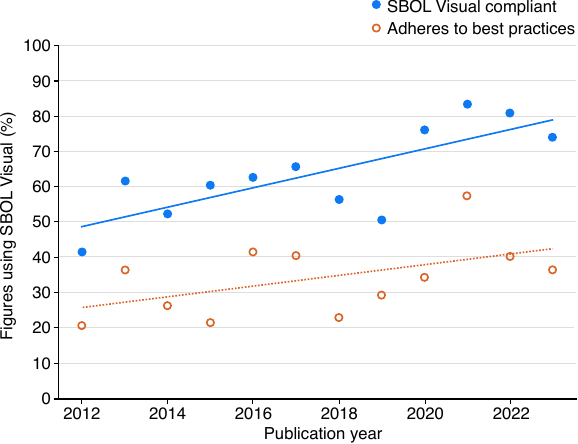}
    \caption{SBOL Visual adoption in genetic diagrams published in ACS Synthetic Biology from 2012 to 2023. Compliance was met if a figure adhered to mandatory rules (blue filled circles), while a figure was said to follow best practices if it adhered to all recommended guidelines (orange unfilled circles). Solid and dashed lines denote linear best-fits for compliance and adherence to best practices, respectively.}
    \label{fig:SBOLVadoption}
\end{figure}

\clearpage
\begin{figure}[p]
    \centering 
	\includegraphics[width=1.0\linewidth]{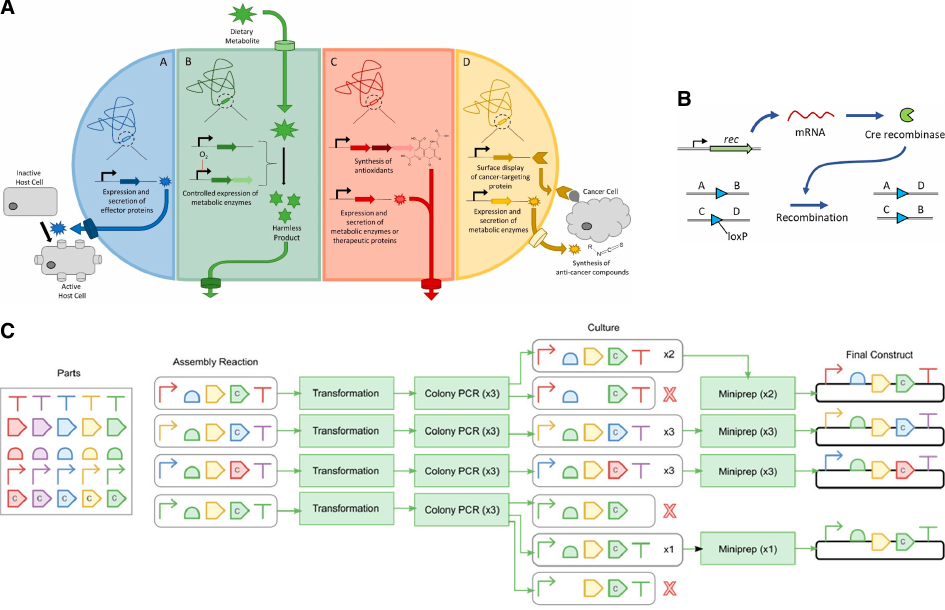}
	\caption{Recent examples of diagrams in ACS Synthetic Biology that make good use of SBOL Visual. (\textsf{\textbf{A}}) Kelly \textit{et al.} explore gene regulatory pathways for strategies to combat metabolic disease and cancer using engineered commensal microbes~\cite{kellyLiving2020}. (\textsf{\textbf{B}}) Okauchi \textit{et al.} develop a Cre recombinase activity assay~\cite{okauchi_cell-free_2021}. (\textsf{\textbf{C}}) Craig \textit{et al.} use combinatorial assembly to create constructs where each is designed to express a specific coding sequence (CDS) using different promoters, ribosome binding sites (RBS), C-terminal epitope tags, and terminator~\cite{craig_LeafLims_2017}.}
    \label{fig:SBOLVexamples}
\end{figure}

\end{document}